\documentclass[superscriptaddress,preprintnumbers,amsmath,amssymb,nofootinbib,eqsecnum,a4paper,secnumarabic,11pt]{revtex4}         
\usepackage[pdftex]{graphicx,color}
\usepackage{ulem}
\usepackage{hyperref}  
\usepackage{titlesec}     
\newcommand*{\justifyheading}{\raggedright}
\titleformat{\chapter}[display]
  {\normalfont\huge\bfseries\justifyheading}{\chaptertitlename\ \thechapter}
  {20pt}{\Huge}
\titleformat{\section}
  {\normalfont\Large\bfseries\justifyheading}{\thesection}{1em}{}
\titleformat{\subsection}
  {\normalfont\large\bfseries\justifyheading}{\thesubsection}{1em}{}
\titleformat{\subsubsection}
  {\normalfont\bfseries\justifyheading}{\thesubsubsection}{1em}{}
\usepackage[english]{babel}

\begin{document}


\title{Muon specific two-Higgs-doublet model}

\author{Tomohiro Abe}
\affiliation{
  Institute for Advanced Research,
  Nagoya University, Furo-cho Chikusa-ku, Nagoya, Aichi, 464-8602 Japan
}
\affiliation{
  Kobayashi-Maskawa Institute for the Origin of Particles and the
  Universe, Nagoya University,
  Furo-cho Chikusa-ku, Nagoya, Aichi, 464-8602 Japan
}

\author{Ryosuke Sato}
\affiliation{Department of Particle Physics and Astrophysics, 
Weizmann Institute of Science, Rehovot 7610001, Israel}

\author{Kei Yagyu}
\affiliation{INFN, Sezione di Firenze, and Department of Physics and Astronomy, 
University of Florence, Via G. Sansone 1, 50019 Sesto Fiorentino, Italy}

\begin{abstract}

We investigate a new type of a two-Higgs-doublet model as a solution of the muon $g-2$ anomaly. 
We impose a softly-broken $Z_4$ symmetry to forbid tree level flavor changing neutral currents in a natural way.
This $Z_4$ symmetry restricts the structure of Yukawa couplings.
As a result, extra Higgs boson couplings to muons are enhanced by a factor of $\tan\beta$,
while their couplings to all the other standard model fermions are suppressed by $\cot\beta$.
Thanks to this coupling property, 
we can avoid the constraint from leptonic $\tau$ decays in contrast to the 
lepton specific two-Higgs-doublet model, 
which can explain the muon $g-2$ within the 2$\sigma$ level but cannot within the $1\sigma$ level due to this constraint.
We find that the model can explain the muon $g-2$ within the 1$\sigma$ level 
satisfying constraints from perturbative unitarity, vacuum stability, electroweak precision measurements, and current LHC data. 

\end{abstract}

\maketitle

\newpage
\section{Introduction}

It is well known that
the measured value of the muon anomalous magnetic moment ($g-2$)~\cite{Bennett:2006fi}
deviates from the standard model (SM) prediction~\cite{Davier:2010nc, Hagiwara:2011af} with more than 3$\sigma$.
This large deviation has been a long standing problem in particle physics,
and many models beyond the SM have been studied to solve this discrepancy~\cite{0902.3360}.
Since the new experiments are planed at Fermilab \cite{Grange:2015fou} and J-PARC \cite{Iinuma:2011zz},
it is worthwhile to find a good benchmark model that solves this problem.

Among various scenarios, the lepton specific (Type-X) two-Higgs-doublet model (THDM) gives a simple solution to explain the muon $g-2$ anomaly~\cite{1409.3199,1412.4874}.\footnote{
Other scenarios of THDMs without the natural flavor conservation~\cite{Glashow:1976nt} are discussed in Refs.~\cite{1502.07824, 1511.05162, 1511.08880}.} 
This model is known as one of the four THDMs~\cite{Barger:1989fj,hep-ph/9401311,0902.4665} with a softly-broken $Z_2$ symmetry
which is imposed to avoid flavor changing neutral currents (FCNCs) at the tree level~\cite{Glashow:1976nt}. 
This model contains additional Higgs bosons, namely a CP-even ($H^0$), a CP-odd ($A^0$), and charged ($H^{\pm}$) Higgs bosons.
Their couplings to the SM charged leptons are enhanced by a factor of $\tan\beta$ which is the ratio of two Vacuum Expectation Values (VEVs) of the two doublet Higgs fields. 
Although this enhancement can significantly reduce the discrepancy in the muon $g-2$, 
its amount is severely constrained from precision measurements of the leptonic $\tau$ decay: $\tau \to \mu \nu_\tau \bar{\nu}_\mu$ whose 
amplitude with the $H^\pm$ mediation is proportional to $\tan^2\beta$. 
Consequently, it turns out difficult to explain the muon $g-2$ anomaly within the 1$\sigma$ level~\cite{1504.07059,1605.06298}. 

In this paper, we propose a new type of the THDM that avoids the constraint from the $\tau$ decay
without losing the advantage of the Type-X THDM.
We impose a softly-broken $Z_4$ symmetry to forbid tree level FCNCs in a natural way as in the Type-X THDM. 
This $Z_4$ symmetry is also important to restrict the structure of Yukawa couplings. 
As a result, only the additional Higgs boson couplings to muons are enhanced by a factor of $\tan\beta$, 
while their couplings to all the other SM fermions are suppressed by $\cot\beta$.
We call this model the ``muon specific THDM ($\mu$THDM)''. 
Thanks to this coupling property, 
the large contribution to the leptonic $\tau$ decay amplitude by $\tan^2\beta$ provided in the Type-X THDM disappears in the $\mu$THDM because of the cancellation of the $\tan\beta$ factor between the tau and the muon Yukawa couplings to $H^\pm$. 
This is a crucial difference of this model from the Type-X THDM. 
We will show that the $\mu$THDM can explain the muon $g-2$ anomaly within the $1\sigma$ level in the parameter space allowed by bounds from 
perturbative unitarity, vacuum stability, electroweak precision measurements, and current LHC data.

This paper is organized as follows. 
After describing our model in Sec.~\ref{sec:model}, 
we discuss constraints on model parameters 
from perturbative unitarity, vacuum stability, electroweak precision measurements, and current LHC data in Sec.~\ref{sec:g2}. 
In addition, we show that 
the parameter space which explains the muon $g-2$ anomaly within $1\sigma$ is allowed by these constraints 
We devote Sec.~\ref{sec:conclusion} for our conclusion.

\section{Model}\label{sec:model}

\subsection{Lagrangian}

The Higgs sector of the $\mu$THDM is composed of two SU(2)$_L$ doublet scalar fields $H_1$ and $H_2$. 
We impose a softly-broken $Z_4$ symmetry to prevent tree level FCNCs. 
The charge assignment for the SM fermions and the Higgs fields are summarized in Table~\ref{tab:matter}.\footnote{Our model can be extended so as to realize non-zero masses of left-handed neutrinos and large mixing angles between $\nu_\mu$ and $\nu_{e,\tau}$ which are
observed by neutrino experiments. 
We discuss such extension without a hard breaking of the $Z_4$ symmetry in Appendix~\ref{app:neutrino}.}
\begin{table}
\caption{Particle contents and the charge assignment.}
\label{tab:matter}
\begin{tabular}{|c||c|c|c||c|c|c||c|c|c||c|c|}
\hline
          & $q_L^{j}$ & $u_R^j$ & $d_R^j$ & $\ell_L^e$ & $\ell_L^\tau$ & $\ell_L^\mu$ & $e_R$ & $\tau_R$ & $\mu_R$ & $H_1$ & $H_2$ \\ \hline
SU(3)$_c$ & 3         & 3       & 3       & 1          & 1             & 1            & 1     & 1        & 1       & 1     & 1     \\ 
SU(2)$_L$ & 2         & 1       & 1       & 2          & 2             & 2            & 1     & 1        & 1       & 2     & 2     \\ 
 U(1)$_Y$ & 1/6       & 2/3     & $-1/3$  & $-1/2$     & $-1/2$        & $-1/2$       & $-1$  & $-1$     & $-1$    & 1/2   & 1/2   \\ 
    Z$_4$ & 1         & 1       & 1       & 1          & 1             & $i$          & 1     & 1        & $i$     & $-1$  & 1     \\ \hline
\end{tabular}
\end{table}

The Yukawa interaction terms under this charge assignment are given by\footnote{
We discuss the possibility of other discrete symmetries which realize this Yukawa structure in Appendix~\ref{sec:Zn}.
}
\begin{align}
{\cal L}^{\text{Yukawa}}
=&
- \bar{q}_L \tilde{H}_2 Y_u u_R
- \bar{q}_L H_2 Y_d d_R
- \bar{L}_L H_1 Y_{\ell 1} E_R
- \bar{L}_L H_2 Y_{\ell 2} E_R
+ (h.c), \label{yuk}
\end{align}
where $\tilde{H}_2 = i\sigma^2 H^{*}_2$, and 
$Y_u$, $Y_d$, $Y_{\ell 1}$ and $Y_{\ell 2}$ are $3 \times 3$ matrices in generation space. 
The left(right)-handed lepton filed $L_L$ $(E_R)$ is defined as 
\begin{align}
L_L = (\ell_L^e ,  \ell_L^{\tau} ,  \ell_L^{\mu} )^T, \quad
E_R = (e_R^{} ,  \tau_R^{} , \mu_R^{})^T. 
\end{align}
The $Z_4$ symmetry restricts the structure of the lepton Yukawa matrices as follows:
\begin{align}
 Y_{\ell 1}
=
\begin{pmatrix}
 0 & 0 & 0 \\
 0 & 0 & 0 \\
 0 & 0 & y_{\mu}
\end{pmatrix}
,
\quad
 Y_{\ell 2}
=
\begin{pmatrix}
 y_{e} & y_{e\tau} & 0 \\
 y_{\tau e} & y_{\tau} & 0 \\
 0 & 0 & 0 
\end{pmatrix}
.
\label{eq:texture_of_leptonYukawa}
\end{align}
We can take $y_{e \tau} = y_{\tau e} = 0$ by field rotations without loss of generality.

The Higgs potential takes the same form as in the THDM with a softly-broken $Z_2$ symmetry:
\begin{align}
 V=&
  m_1^2 H_1^{\dagger} H_1
+ m_2^2 H_2^{\dagger} H_2
- \left(m_3^2 H_1^{\dagger} H_2 + (h.c.) \right)
+ \frac{ \lambda_1}{2} (H_1^{\dagger} H_1)^2
+ \frac{\lambda_2}{2}  (H_2^{\dagger} H_2)^2
\notag \\
&
+ \lambda_3 (H_1^{\dagger} H_1) (H_2^{\dagger} H_2)
+ \lambda_4 (H_1^{\dagger} H_2) (H_2^{\dagger} H_1)
+
\left(
  \frac{1}{2} \lambda_5 (H_1^{\dagger} H_2)^2 
+ (h.c.)
\right),
\label{eq:HiggsPotential}
\end{align}
where $m_1^2$, $m_2^2$, $\lambda_1$, $\lambda_2$, $\lambda_3$, and $\lambda_4$ are real.
In general, $m_3^2$ and $\lambda_5$ are complex, but we assume these two parameters to be real for simplicity, 
by which the Higgs potential is CP-invariant.

We parametrize the component fields of the Higgs doublets by
\begin{align}
H_i
=&
\left(
\begin{matrix}
 \pi^{+}_i 
\\ 
 \frac{1}{\sqrt{2}}
\left(
v_i
+
\sigma_i
-
i \pi_i^{3}
\right)
\end{matrix}
\right)
,
\quad \quad
(i = 1, 2),  \label{components}
\end{align}
where $v_1~(v_2)$ is the VEV of the $H_1~(H_2)$ field.
It is convenient to express these two VEVs in terms of $v$ and $\tan\beta$ defined by 
$v \equiv \sqrt{v_1^2 + v_2^2} \simeq (\sqrt{2}G_F)^{-1/2}\simeq 246$ GeV with $G_F$ being the Fermi constant and $\tan\beta \equiv v_2/v_1$, respectively.\footnote{The exact relation between $v$ and $G_F$ is given in Appendix \ref{sec:EWPTapp}.} 
The mass eigenstates of the scalar bosons and their relation to the gauge eigenstates expressed in Eq.~(\ref{components}) 
are given by the following rotations:
\begin{align}
\left(
\begin{matrix}
 \pi_{Z} \\
 A^0 \\
\end{matrix}
\right)
=&
\left(
\begin{matrix}
 \cos\beta     &    \sin\beta \\
 -\sin\beta     &   \cos\beta 
\end{matrix}
\right)
\left(
\begin{matrix}
 \pi_1^{3} \\
 \pi_2^{3}
\end{matrix}
\right)
,
\\
\left(
\begin{matrix}
 \pi_{W^{\pm}} \\
 H^{\pm} \\
\end{matrix}
\right)
=&
\left(
\begin{matrix}
 \cos\beta     &    \sin\beta \\
 -\sin\beta     &   \cos\beta 
\end{matrix}
\right)
\left(
\begin{matrix}
 \pi_1^{\pm} \\
 \pi_2^{\pm}
\end{matrix}
\right)
,
\\
\left(
\begin{matrix}
 H^0 \\
 h \\
\end{matrix}
\right) 
=&
\left(
\begin{matrix}
 \cos \alpha
& \sin \alpha
\\ 
 -\sin \alpha
& \cos \alpha
\end{matrix}
\right)
\left(
\begin{matrix}
 \sigma_1 \\
 \sigma_2 \\
\end{matrix}
\right) 
, \label{alpha}
\end{align}
where $\pi_{W^\pm}^{}$ and $\pi_Z^{}$ are the Nambu-Goldstone bosons which are absorbed into the longitudinal component of the $W^\pm$ and $Z$ bosons, respectively. 
We identify $h$ as the discovered Higgs boson with a mass of 125~GeV at the LHC. 
The mixing angle $\alpha$ is expressed by the potential parameters as
\begin{align}
\tan2\alpha = \frac{2(v^2\lambda_{345}-M^2)\tan\beta}{v^2(\lambda_1 - \lambda_2\tan^2\beta) - M^2(1-\tan^2\beta)}, 
\end{align}
where
\begin{align}
 \lambda_{345} \equiv& \lambda_3 + \lambda_4 + \lambda_5, \quad
 M^2 \equiv  \frac{1 + t_\beta^2 }{t_\beta} m_3^2. 
\label{eq:def_M}
\end{align}
The CP-conserving Higgs potential can then be described by the following 8 independent parameters:
\begin{align}
m_{H^\pm}^{},~m_A^{},~m_H^{},~m_h,~M^2,~\alpha,~\beta,~v, \label{para}
\end{align}
where $m_{H^\pm}^{},~m_A^{},~m_H^{}$ and $m_h$ denote the masses of $H^\pm$, $A^0$, $H^0$ and $h$, respectively. 

We introduce the following shorthand notations for the later convenience.
\begin{align}
 s_x = \sin x, \quad c_x = \cos x, \quad t_x = \tan x.
\end{align}

\subsection{Yukawa couplings in large $\tan\beta$ regime}

From Eq.~(\ref{yuk}),  we can extract 
interaction terms for the mass eigenstates of the Higgs bosons with the third generation fermions and the muon as follows:
\begin{align}
{\mathcal L}_\text{int} 
&=
-\sum_{f=t,b,\tau}\frac{m_f}{v}\left[ \left(s_{\beta-\alpha}+\frac{c_{\beta-\alpha}}{t_\beta}\right){\overline f}fh
+\left(c_{\beta-\alpha} - \frac{s_{\beta-\alpha}}{t_\beta}\right){\overline f}fH^0
+2i\frac{I_f}{t_\beta}{\overline f}\gamma_5fA^0\right]\notag\\
&
-\frac{m_\mu}{v}\left[ \left(s_{\beta-\alpha}-t_\beta c_{\beta-\alpha}\right){\overline \mu}\mu h
+\left(c_{\beta-\alpha} +t_\beta s_{\beta-\alpha} \right){\overline \mu}\mu H^0
+it_\beta {\overline \mu}\gamma_5 \mu A^0\right]\notag\\
&-\frac{\sqrt{2}}{v}  \left\{\frac{1}{t_\beta}\left[\overline{t}
\left( m_b \text{P}_R-m_t\, \text{P}_L\right)b\,H^+ + m_{\tau} \overline{\nu_\tau} \, P_R \, \tau \,H^+ \right]
-t_\beta m_{\mu} \overline{\nu_\mu} \, P_R \, \mu \,H^+ +(h.c.) \right\},  \label{int}
\end{align}
where $P_{L}(P_{R})$ is the projection operator for left(right)-handed fermions and $I_f=+1/2 \, (-1/2)$ for $f = t \, (b,\tau,\mu)$. 
The masses of fermions are given by 
\begin{align}
m_\mu = \frac{v}{\sqrt{2}}\frac{y_\mu}{\sqrt{1+t_\beta^2}},~~
m_f = \frac{v}{\sqrt{2}}\frac{y_f t_\beta}{\sqrt{1+t_\beta^2}}~~~(f = t,b,\tau),  \label{mass}
\end{align}
From Eq.~(\ref{int}), it is clear that only the muon couplings to the extra Higgs bosons are enhanced by taking large $\tan\beta$. 

In order to solve the muon $g-2$ anomaly, we need a large value of $\tan\beta$ to obtain significant loop effects of extra Higgs bosons as  
we will show it in the next section. 
Let us here discuss how large value of $\tan\beta$ we can take without spoiling perturbativity. 
From Eq.~(\ref{mass}) we obtain 
\begin{align}
 y_\mu = \frac{\sqrt{2}m_\mu}{v} \sqrt{1 + t^2_\beta}  \simeq 0.6 \left(\frac{t_\beta}{1000} \right).
\end{align}
For example, $t_\beta \lesssim 5000$ for $y_{\mu} \lesssim 3$.
Clearly from Eq.~(\ref{mass}), all the other Yukawa couplings approach to the corresponding SM value in large $\tan\beta$, 
so that they do not cause the violation of perturbativity in this limit.

\subsection{Scalar quartic couplings in large $\tan\beta$ regime}

Next, we discuss the behavior of the Higgs quartic couplings in the large $\tan\beta$ regime.
All these couplings (times $v^2$) can be rewritten in terms of the parameters shown in Eq.~(\ref{para}) as
\begin{align}
 \lambda_1 v^2
=&
\left(
m_h^2 c_{\beta - \alpha}^2 + m_H^2 s_{\beta - \alpha}^2- M^2\right)t_\beta^2
+
2(m_H^2 - m_h^2)s_{\beta - \alpha} c_{\beta - \alpha}\, t_\beta
+ m_h^2 s_{\beta - \alpha}^2  +  m_H^2 c_{\beta - \alpha}^2 , \label{lam1}  \\
 \lambda_2 v^2
=& m_h^2 s_{\beta - \alpha}^2 + m_H^2 c_{\beta - \alpha}^2
-  2(m_H^2 - m_h^2)   \frac{c_{\beta - \alpha} s_{\beta - \alpha}}{t_\beta}
+( m_h^2 c_{\beta - \alpha}^2 + m_H^2 s_{\beta - \alpha}^2- M^2 )\frac{1}{t_\beta^2}
,\\
 \lambda_3 v^2
 =&(m_H^2 - m_h^2) c_{\beta - \alpha} s_{\beta - \alpha} \left( t_\beta - \frac{1}{t_\beta} \right)
+ 2 m_{H^{\pm}}^2 - M^2 + m_H^2 - m_h^2(c_{\beta - \alpha}^2 -s_{\beta - \alpha}^2 ) ,\\
 \lambda_{4}v^2
=& M^2 - m_A^2 - 2 (m_{H^{\pm}}^2 - m_A^2)  ,\\
 \lambda_{5}v^2
=& M^2 - m_A^2 . \label{lam5}
\end{align}
We find that in the large $\tan\beta$ regime, $\lambda_1$ and $\lambda_3$ can be very large because they are proportional to $t_\beta^2$ and $t_\beta$, respectively, which 
causes the validity of perturbative calculations to be lost. 
In order to keep $\lambda_1$ and $\lambda_3$ to be reasonable values, we can 
take $M^2$ and $s_{\beta - \alpha}$ so as to cancel the large contribution from the $t_\beta^2$ and $t_\beta$ terms as follows:
\begin{align}
 M^2
& =  m_h^2 c_{\beta - \alpha}^2
+ m_H^2 s_{\beta - \alpha}^2
- 2 s_{\beta - \alpha} c_{\beta - \alpha} ( m_h^2 - m_H^2) \frac{1}{t_\beta}
- X v^2 \frac{1}{t_\beta^2}
, \label{eq:adjust_M2}\\
s_{\beta - \alpha} 
& =
1
, \label{eq:adjust_sBA}
\end{align}
where $X$ is an arbitrary number.

It is worth noting that in the limit $s_{\beta - \alpha} \to 1$ (the so-called alignment limit~\cite{Gunion:2002zf}), 
the SM-like Higgs boson $h$ couplings to weak bosons $g_{hVV}^{}$ $(V=W,Z)$ and fermions $g_{hff}$ become the same value as those of the SM Higgs boson at the tree level, because 
these are given by $g_{hVV}^{} = g_{hVV}^{\text{SM}} s_{\beta - \alpha}$, $g_{hff}^{} = g_{hff}^{\text{SM}} (s_{\beta - \alpha} + c_{\beta-\alpha}/t_\beta)$ ($f \neq \mu $)
and $g_{h\mu\mu}^{} = g_{h\mu\mu}^{\text{SM}} (s_{\beta - \alpha} - c_{\beta-\alpha}t_\beta)$. 
Because no large deviation in the Higgs boson couplings from the SM prediction has been discovered at current LHC data~\cite{1606.02266}, 
our choice $s_{\beta - \alpha} =1$ is consistent with these results.
After imposing Eqs.~\eqref{eq:adjust_M2} and \eqref{eq:adjust_sBA}, we find
\begin{align}
 \lambda_1 
=&
\frac{m_h^2}{v^2}  +  X
,\\
 \lambda_2
=&
\frac{m_h^2}{v^2} + \frac{X}{t_\beta^4},\\
 \lambda_{3}
=&
\frac{2 m_{H^{\pm}}^2 -2m_H^2 + m_h^2}{v^2} + \frac{X}{t_\beta^2}
,\\
 \lambda_{4}
=&
\frac{m_H^2 + m_A^2 - 2m_{H^{\pm}}^2}{v^2} - \frac{X}{t_\beta^2}
,\\
 \lambda_{5}
=&
\frac{m_H^2 - m_A^2}{v^2}- \frac{X}{t_\beta^2}.    \label{lam55}
\end{align}
These $\lambda$'s are at most ${\cal O}(1)$ as long as we take $m_{H^\pm}^2 \sim m_{A}^2 \sim m_H^2$, so that  we can still treat them as perturbation.
We take $X = 0$ for simplicity in the following analysis.
Constraints from perturbative unitarity is discussed in Sec.~\ref{sec:PU}.

\section{Muon $g-2$ and Constraints on parameter space \label{sec:g2}}

In this section, we discuss the muon $g-2$ anomaly and various constraints on the model parameters. 

\subsection{Muon $g-2$}

In the scenario with $s_{\beta-\alpha} = 1$ as discussed in the previous section, 
new contributions to $a_\mu \equiv (g-2)/2$ purely comes from the loop contributions of $H^0$, $A^0$ and $H^\pm$, because 
the couplings of $h$ becomes exactly the same as those of the SM Higgs boson at the tree level. 
One-loop diagram contributions to $\delta a_\mu \equiv a_\mu - a_\mu^{\text{SM}}$ from additional Higgs boson loops are calculated as~\cite{Dedes:2001nx}  
\begin{align}
 \delta a_\mu^H &= \frac{G_F m_\mu^2}{4\sqrt{2}\pi^2} t^2_\beta \left( \frac{c_{\beta - \alpha}}{t_\beta} + s_{\beta - \alpha} \right)^2 r_H^{} f_H(r_H^{}),\\
 \delta a_\mu^A &= \frac{G_F m_\mu^2}{4\sqrt{2}\pi^2} t^2_\beta r_A f_A(r_A),\\
 \delta a_\mu^{H^\pm} &= \frac{G_F m_\mu^2}{4\sqrt{2}\pi^2} t_\beta^2 r_{H^\pm} f_{H^\pm}(r_{H^\pm}),
\end{align}
where $r_{H,A,H^\pm} = m_\mu^2 / m_{H,A,H^\pm}^2$ and
\begin{align}
  f_H(r) &= \int_0^1 dx \frac{x^2(2-x)}{r x^2 - x + 1}, \\
  f_A(r) &= \int_0^1 dx \frac{-x^3}{r x^2 - x + 1}, \\
  f_{H^\pm}(r) &= \int_0^1 dx \frac{-x^2(1-x)}{r x^2 + (1-r) x}.
\end{align}
For $r_{H,A,H^\pm} \ll 1$, we can approximate the above formulae as follows:
\begin{align}
 \delta a_\mu^H 
&\simeq  \frac{G_F m_\mu^2}{4\sqrt{2}\pi^2}t_\beta^2
\left(s_{\beta - \alpha} + \frac{c_{\beta - \alpha}}{t_\beta} \right)^2
 \frac{m_\mu^2}{m_H^2} 
 \left( - \frac{7}{6} - \ln  \frac{m_\mu^2}{m_H^2} \right)
,\\
 \delta a_\mu^A 
&\simeq  \frac{G_F m_\mu^2}{4\sqrt{2}\pi^2}
 \frac{m_\mu^2}{m_A^2} 
 \left( \frac{11}{6} + \ln  \frac{m_\mu^2}{m_A^2} \right)
,\\
 \delta a_\mu^{H^\pm} 
&\simeq  
\frac{G_F m_\mu^2}{4\sqrt{2}\pi^2}
 \frac{m_\mu^2}{m_{H^\pm}^2} 
 \left( -\frac{1}{6} \right).
\end{align}

We here briefly mention the contribution from two-loop Barr-Zee diagrams~\cite{BZ1,BZ2}. 
In the Type-II and Type-X THDMs, the Barr-Zee diagrams also give important contributions, 
because the tau and/or bottom Yukawa couplings to the additional Higgs bosons
can be enhanced by $\tan\beta$. 
As a result, these two-loop contributions can be comparable to the one-loop diagram. 
However, in the present model, the both tau and bottom Yukawa couplings are suppressed by $\cot\beta$ as seen in Eq.~(\ref{int}). 
Therefore, the contribution from two-loop diagrams is simply suppressed by the loop factor, so that these cannot be important.  
We thus only consider the one-loop diagram for the muon $g-2$.

\begin{figure}[tb]
 \includegraphics[width=0.31\hsize]{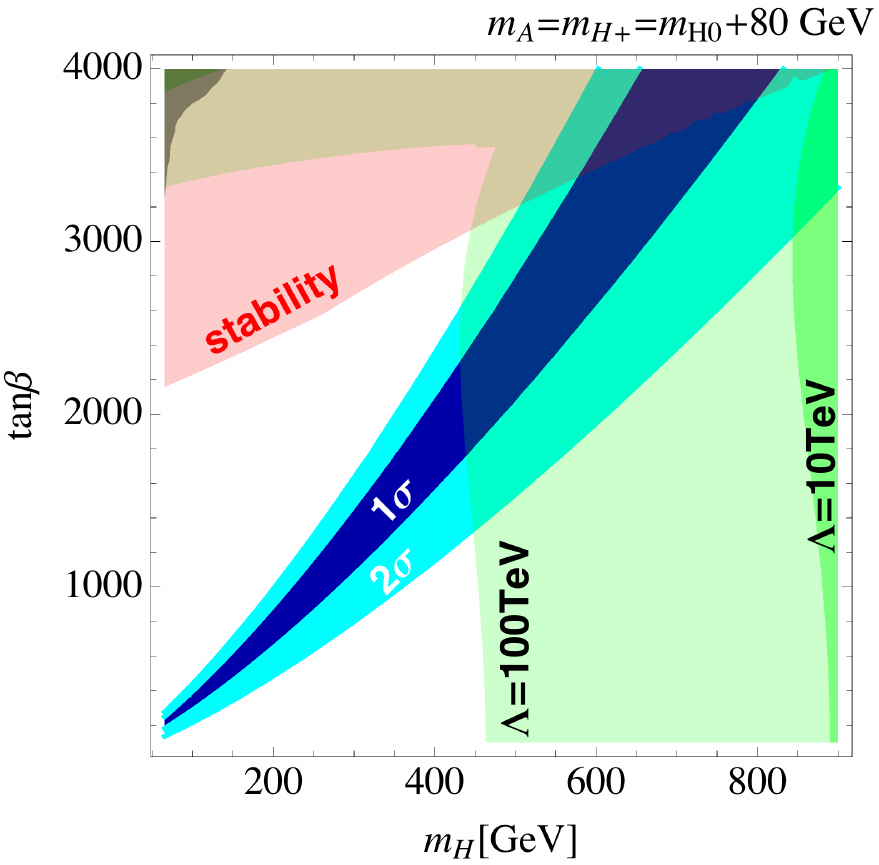} \quad
 \includegraphics[width=0.31\hsize]{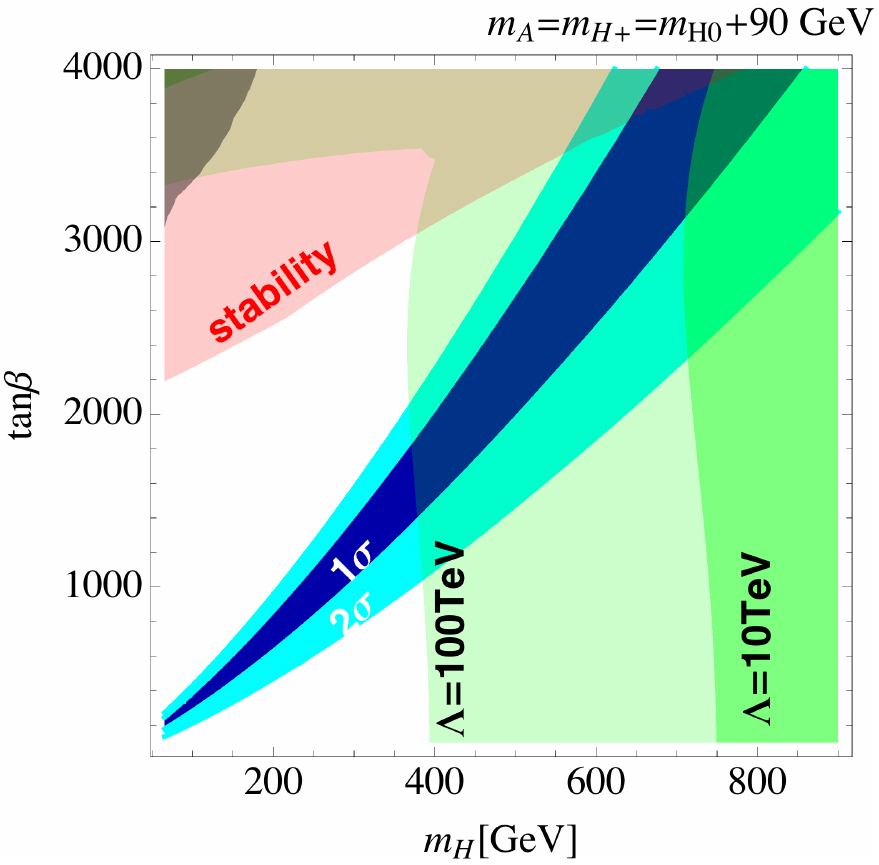} \quad
 \includegraphics[width=0.31\hsize]{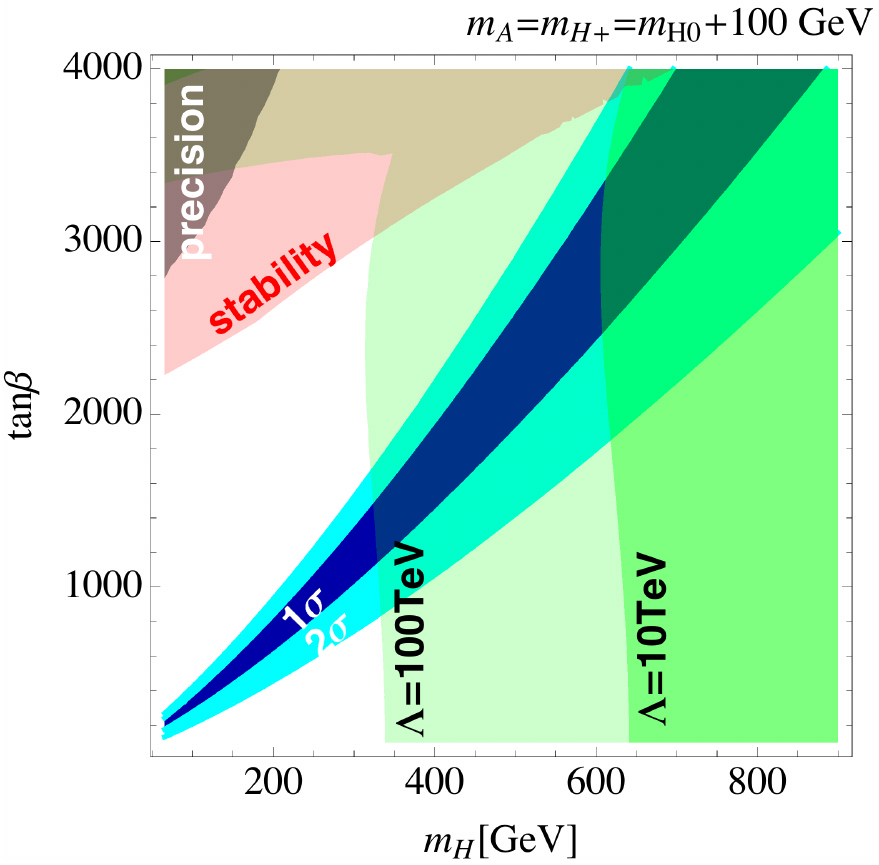} \quad
\caption{
Regions where the prediction for the muon $g-2$ is consistent with the measurement 
within 1$\sigma$ (blue) and 2$\sigma$ (cyan).
The green (darker green) region shows the cutoff scale to be less than 100 (10)~TeV given 
by the perturbative unitarity bound (see Sec.~\ref{sec:PU}). 
The red region indicates the cutoff scale to be less than 10~TeV given 
by the vacuum stability bound (see Sec.~\ref{sec:PU}). 
The gray region is excluded by the electroweak precision measurements at 95\% CL (see Sec.~\ref{sec:ewpt}). 
}
\label{fig:g-2_and_constraints}
\end{figure}

Numerical results for $\delta a_\mu$ are shown in Fig.~\ref{fig:g-2_and_constraints} on the $m_H^{}$--$\tan\beta$ plane.
The blue and cyan regions show the regions of parameter space where we can explain the muon $g-2$ within 1$\sigma$ and 2$\sigma$, respectively.
Here, we consider the case with $H^{0}$ to be the lightest of all the additional Higgs bosons, and we display the three cases for the 
mass difference between $m_H^{}$ and $m_A^{}(=m_{H^\pm}^{})$ being 80 (left), 90 (center) and 100 (right) GeV. 
We can see that the prediction of $\delta a_\mu$ is not changed so much among these three cases. 
We find that 
the discrepancy of the muon $g-2$ becomes 1$\sigma$ by taking, e.g., $m_H^{}=300 (600)$ GeV with $\tan\beta=1000\,(3000)$. 

\subsection{Constraints on scalar quartic couplings }
\label{sec:PU}

The scalar quartic couplings $\lambda_1$--$\lambda_5$ in the Higgs potential can be constrained 
by taking into account the following theoretical arguments. 
Such constraint can be translated into the bound on the physical Higgs boson masses and mixing angles via Eqs.~(\ref{lam1})--(\ref{lam5}). 

First, the Higgs potential must be bounded from below in any direction of the scalar field space. 
The sufficient condition to guarantee the vacuum stability is given by~\cite{PHRVA.D18.2574,Sher:1988mj,PHLTA.B449.89,PHLTA.B471.182}
\begin{align}
\lambda_1>0, \quad \lambda_2>0,
\quad \sqrt{\lambda_1\lambda_2}+\lambda_3+\text{MIN}(0,\lambda_4+\lambda_5,\lambda_4-\lambda_5)>0.
\label{eq:stability}
\end{align}

Next, perturbative unitarity requires that 
$s$-wave amplitude matrices for elastic scatterings of scalar boson 2-body to 2-body processes
must not be too large to satisfy S matrix unitarity. 
This perturbative unitarity condition is expressed as 
\begin{align}
|a_{i,\pm}^0|\leq\frac{1}{2}, \label{pv1}
\end{align}
where $a_{i,\pm}^0$ are the eigenvalues of such $s$-wave amplitude matrices. 
In the CP-conserving THDMs, these eigenvalues are given by~\cite{PHLTA.B313.155,hep-ph/0006035,PHRVA.D72.115010,Kanemura:2015ska}:
\begin{align}
a_{1,\pm}^0 &=  \frac{1}{32\pi}
\left[3(\lambda_1+\lambda_2)\pm\sqrt{9(\lambda_1-\lambda_2)^2+4(2\lambda_3+\lambda_4)^2}\right],\\
a_{2,\pm}^0 &=
\frac{1}{32\pi}\left[(\lambda_1+\lambda_2)\pm\sqrt{(\lambda_1-\lambda_2)^2+4\lambda_4^2}\right],\\
a_{3,\pm}^0 &= \frac{1}{32\pi}\left[(\lambda_1+\lambda_2)\pm\sqrt{(\lambda_1-\lambda_2)^2 + 4\lambda_5^2}
\right],\\
a_{4,\pm}^0 &= \frac{1}{16\pi}(\lambda_3+2\lambda_4 \pm \lambda_5),\\
a_{5,\pm}^0 &= \frac{1}{16\pi}(\lambda_3\pm\lambda_4),\\
a_{6,\pm}^0 &= \frac{1}{16\pi} (\lambda_3 \pm \lambda_5).  \label{pv2}
\end{align}

We impose the above two conditions given in Eqs.~(\ref{eq:stability}) and (\ref{pv1}) at an arbitrary energy scale $\mu$. 
In this case, all the scalar quartic couplings $\lambda_1$--$\lambda_5$ should be understood as a function of $\mu$, where their energy dependence 
are determined by solving renormalization group equations. 
In addition, we require that no Landau pole appears up to a certain energy scale, and we call this the triviality bound. 
From the above consideration, 
we can define the cutoff scale of the theory $\Lambda_{\text{cutoff}}$ in such a way that 
one of the three conditions, i.e., the perturbative unitarity, the vacuum stability and the triviality bounds is not satisfied.  
The renormalization group equations are expressed by a set of $\beta$-functions for dimensionless parameters defined by 
\begin{align}
 \mu \frac{d}{d\mu} c = \frac{1}{(4\pi)^2} \beta_{c}. 
\end{align}
We calculate the $\beta$-functions by using \texttt{SARAH}~\cite{1309.7223}.
They are approximately given as follows:
{\allowdisplaybreaks  \begin{align} 
\beta_{g_1} \simeq& 7 g_{1}^{3}, \\ 
\beta_{g_2} \simeq& -3 g_{2}^{3}, \\ 
\beta_{g_3} \simeq& -7 g_{3}^{3}, \\ 
\beta_{\lambda_1}  \simeq&
+\frac{3}{4} g_{1}^{4} +\frac{3}{2} g_{1}^{2} g_{2}^{2} +\frac{9}{4} g_{2}^{4} -3 g_{1}^{2} \lambda_1 -9 g_{2}^{2} \lambda_1 +12 \lambda_{1}^{2} +4 \lambda_{3}^{2} +4 \lambda_3 \lambda_4 +2 \lambda_{4}^{2} +2 \lambda_{5}^{2} \nonumber \\ 
 &+4 \lambda_1 y_\mu^2  -4 y_\mu^4, 
\\
\beta_{\lambda_2}  \simeq&  
+\frac{3}{4} g_{1}^{4} +\frac{3}{2} g_{1}^{2} g_{2}^{2} +\frac{9}{4} g_{2}^{4} -3 g_{1}^{2} \lambda_2 -9 g_{2}^{2} \lambda_2 +12 \lambda_{2}^{2} +4 \lambda_{3}^{2} +4 \lambda_3 \lambda_4 +2 \lambda_{4}^{2} +2 \lambda_{5}^{2} \nonumber \\ 
 &  +12 \lambda_2 y_t^2 -12  y_t^4, 
\\ 
\beta_{\lambda_3} \simeq&  
 \lambda_3 \left(
2 y_\mu^2 +6 y_t^2
-3 g_{1}^{2}  -9 g_{2}^{2} +6 \lambda_1  +6 \lambda_2  +4 \lambda_{3}
\right)
\nonumber\\
&
+\frac{3}{4} g_{1}^{4} -\frac{3}{2} g_{1}^{2} g_{2}^{2} +\frac{9}{4} g_{2}^{4} 
+2 \lambda_1 \lambda_4 +2 \lambda_2 \lambda_4 +2 \lambda_{4}^{2} +2 \lambda_{5}^{2} 
,
\\
\beta_{\lambda_4}  \simeq &
3 g_{1}^{2} g_{2}^{2}
+8 \lambda_{5}^{2} 
+ \lambda_4 \left( 2 \lambda_1  +2 \lambda_2 +8 \lambda_3  +4 \lambda_{4}  -3 g_{1}^{2}  -9 g_{2}^{2}  +2 y_\mu^2 +6 y_t^2 \right)
,\\
\beta_{\lambda_5}  \simeq  &
\lambda_5
\left(
2 \lambda_1  +2 \lambda_2  +8 \lambda_3  +12 \lambda_4  
-3 g_{1}^{2}  -9 g_{2}^{2} 
+2  y_\mu^2 +6  y_t^2
\right),
\\ 
\beta_{y_t} \simeq&
\frac{9}{2} y_t^3 
 +y_t \Big(  -8 g_{3}^{2}  -\frac{17}{12} g_{1}^{2}  -\frac{9}{4} g_{2}^{2}  \Big),
\\
\beta_{y_\mu} \simeq& 
\frac{5}{2} y_\mu^3 - \frac{3}{4} y_\mu \Big(5 g_{1}^{2}  + 3 g_{2}^{2} \Big).
\end{align}} 
Here, we take into account the $y_t$ and $y_\mu$ dependence, and all the other Yukawa couplings are neglected because of their smallness. 
In addition, we ignore higher loop contributions.

In Fig.~\ref{fig:g-2_and_constraints}, we show the $\Lambda_{\text{cutoff}}$ dependence on the $m_H^{}$--$\tan\beta$ plane.  
The regions filled by green (darker green) indicate those with $\Lambda_{\text{cutoff}}\leq 100$ (10) TeV due to the perturbative unitarity bound or 
the triviality bound. 
In addition, the regions filled by red show those with $\Lambda_{\text{cutoff}}\leq 10$ TeV due to the vacuum stability condition. 
If we assume that the model is valid up to 10~TeV and explains the muon $g-2$ within 1$\sigma$,
then the mass of $H^0$ should be smaller than 800~GeV.

\subsection{Constraints from the electroweak precision measurements \label{sec:ewpt}}

The oblique $S$, $T$ and $U$ parameters introduced by Peskin and Takeuchi~\cite{PRLTA.65.964,PHRVA.D46.381}
provide a convenient formalism to discuss the constraint on model parameters from electroweak precision measurements. 
However, we cannot simply apply this formalism to our model, 
because those parameters are formulated under the assumption
that new particles do not give sizable direct corrections to light fermion (including the muon) scattering processes $f_1\bar{f}_2 \to f_3\bar{f}_4$ through vertex corrections and wave function renormalizations. 
The other assumption is that the new physics scale is sufficiently higher than the electroweak scale. 
In our setup, both of them cannot be justified. 
Hence we need to modify the formulation with the $S$, $T$ and $U$ parameters by taking into
account vertex corrections and wave function renormalizations.

By varying the four model parameters ($m_H$, $m_A$, $m_{H^{\pm}}$, $t_\beta$), 
we find that the minimum value of $\chi^2$ to be $\chi_{\text{min.}}^2 = 23.7587$ which is given at  
($m_H$, $m_A$, $m_{H^{\pm}}$, $t_\beta$) = (59.4~GeV, 398~GeV, 402~GeV, 686).
We calculate $\Delta \chi^2 \equiv \chi^2 - \chi_{\text{min.}}^2$ 
by varying $m_H$ and $t_\beta$ with fixed values for $m_A$ and $m_{H^{\pm}}$.
The result is shown in Fig.~\ref{fig:g-2_and_constraints} where the gray region is excluded at 95\% CL.
The detail of our analysis is given in Appendix \ref{sec:EWPTapp}.

\subsection{Constraints and signatures at the LHC experiment\label{sec:LHC}}

Finally, we discuss the constraint on parameters from current LHC data. 

In our model, the quark Yukawa couplings to the additional Higgs bosons are highly suppressed by $\cot\beta$ in the large $\tan\beta$ regime. 
Therefore, the additional neutral Higgs bosons $A^0$ and $H^0$ cannot be produced via the gluon fusion process: $gg \to A^0/H^0$. 
For the same reason, the $gb \to tH^-$ process for the $H^\pm$ production also does not work. 
Moreover, the vector boson fusion process: $qQ \to q'Q'H^0$ is negligible, because the $H^0VV$ couplings are proportional to $c_{\beta-\alpha}$. 
As a result, the main production mode for these Higgs bosons is their pair productions via the $s$-channel mediation of a virtual gauge boson: 
\begin{align}
pp \to Z^* \to H^0 A^0,~~pp \to W^* \to H^\pm A^0/H^\pm H^0,~~pp \to \gamma^*/Z^* \to H^+ H^-. \label{prod}
\end{align}

Because of the muon specific property, the decay branching ratios for $H^{0}$, $A^0$, and $H^\pm$ with the parameter choice in Fig.~\ref{fig:g-2_and_constraints} 
are given as follows:
\begin{align}
 \text{Br}(H^{0} &\to \mu \bar{\mu}) \simeq 1, \\
 \text{Br}(A^{0} &\to \mu \bar{\mu}) + \text{Br}(A^{0} \to H^0 Z) \simeq 1,\\
 \text{Br}(H^{-} &\to \mu \bar{\nu}_{\mu}) + \text{Br}(H^{-} \to H^0 W^{-}) \simeq 1.
\end{align}
The relative magnitude between the above two branching ratios of $A^0$ and that of $H^\pm$ 
mainly depends on the values of $t_\beta$ and the mass difference between $m_H^{}$ and $m_{H^\pm}^{}$. 
For example, we obtain $\text{Br}(A^{0} \to \mu \bar{\mu})$ and 
$\text{Br}(H^{-} \to \mu \bar{\nu}_{\mu})$ to be about 89(99.1)\% and 96(99.7)\%
for $m_{H}^{}=300(600)$ GeV, $m_{H^\pm}^{}- m_H^{} = 100$ GeV and $t_\beta = 1000(3000)$, respectively. 
Therefore, the collider signature of the model is multi-muon final states.

\begin{table}
\centering
\caption{The parameter points that we investigate.
$\sigma_{13\text{TeV}}^{}$ is defined in Eq.~(\ref{13tev}). 
$N_{\mu\text{-THDM}}$ is the expected signal event numbers in the last bin of Fig.~2(b) in Ref.~\cite{CMS:2017wua}. 
${\cal L}_{3\sigma}$ is the integrated luminosity at which we can expect 3$\sigma$ deviation from the SM prediction if we apply the same analysis as Ref.~\cite{CMS:2017wua}.
The data points  with ``-'' in the last column are already excluded. 
}
\label{tab:LHC}
\begin{tabular}{|c|c|c||c|c|c|} 
\hline $m_{H^0}$ [GeV] & $m_{A^{0}}(=m_{H^{\pm}})$ [GeV] & $\tan\beta$ & $\sigma_{13\text{TeV}}$ [fb] 
& $N_{\mu\text{-THDM}}$ & ${\cal L}_{3\sigma}$ [fb$^{-1}$]
\\ \hline
 600 & 700 & 3000 & 0.41   & 6.6 & -\\
 620 & 710 & 3000 & 0.369  & 5.9 & -\\
 640 & 730 & 3100 & 0.316  & 5.2 & 44\\
 660 & 750 & 3300 & 0.2707 & 4.5 & 58\\
 680 & 770 & 3400 & 0.2334 & 3.9 & 75\\
 700 & 790 & 3700 & 0.20   & 3.4 & 97\\
\hline
\end{tabular}
\end{table}

We show the production cross sections in some parameter points given in Table~\ref{tab:LHC}.
Here, the production cross section is defined as the sum of all the modes given in Eq.~(\ref{prod}) at 13~TeV, 
\begin{align}
 \sigma_{13\text{TeV}} 
\equiv&
  \sum_{X = A^0, H^{\pm}} \sigma(pp \to H^0 X)
+ \sum_{Y = H^{\pm}} \sigma(pp \to A^0 Y)
+ \sigma(pp \to H^+ H^-).\label{13tev}
\end{align}
We generate \texttt{UFO} files~\cite{1108.2040} by using \texttt{FeynRules 2.3.3}~\cite{1310.1921}, 
and use \texttt{MadGraph 5}~\cite{1106.0522} to estimate the production cross sections.
Signal events are simulated by using \texttt{MadGraph 5}, \texttt{PYTHIA 6.428}~\cite{hep-ph/0603175}, and \texttt{DELPHES 3.3.3}~\cite{1307.6346}.
We compare the number of events predicted in our model with that of the CMS result for the multi-lepton signal search at 13~TeV with 35.9~fb$^{-1}$ data~\cite{CMS:2017wua}.
We find the last bin of Fig.~2(b) in Ref.~\cite{CMS:2017wua} gives the stringent bound on the mass of $H^0$
because our model predicts three-muon final states with large $p_T$, e.g., via $pp \to H^0H^\pm \to \mu^+\mu^-\mu^\pm\nu$. 
The observed (expected) background event number in the bin is 3(3.5). 
The expected signal event numbers in several parameter points are shown in Table~\ref{tab:LHC}.
We use the the CLs method~\cite{hep-ex/9902006,Read:2000ru,Read:2002hq}, 
and find that the region with $m_{H}^{} \lesssim 640$ GeV is excluded at 95\% CL.
Also, we show the integrated luminosity which is required to give the 3$\sigma$ deviation from the SM expectation for each parameter point.
We can see that the allowed parameter points ($m_{H}^{} \geq 640$ GeV) could give the 3$\sigma$ deviation during the LHC Run 2 experiment.

\section{Conclusions \label{sec:conclusion}}

We have investigated a new type of the THDM, i.e. $\mu$THDM, as a solution of the muon $g-2$ anomaly.
Differently from the other THDMs with a softly-broken $Z_2$ symmetry,
this model predicts that only the muon couplings to the additional Higgs bosons are enhanced by $\tan\beta$, 
while all the other SM fermion couplings to them are suppressed by $\cot\beta$. 
Thanks to this coupling property, the $\mu$THDM can avoid the strong constraint from the leptonic $\tau$ decay
in contrast to the Type-X THDM which cannot explain the muon $g-2$ within the $1\sigma$ level due to this constraint.
We find that the $\mu$THDM can explain the muon $g-2$ within the 1$\sigma$ level 
satisfying constraints from perturbative unitarity, vacuum stability, electroweak precision measurements, and current LHC data. 

We have found that large $\tan\beta$ is required to solve the muon $g-2$ anomaly within the $1\sigma$ level. 
Its typical values is ${\cal O}(1000)$ with the masses of the additional Higgs bosons to be in the range of 100--1000 GeV. 
The large $\tan\beta$ is equivalent to the large muon Yukawa coupling, $y_{\mu} \sim {\cal O}(1)$.
In order to see the effect of such large Yukawa coupling,
we have studied the constraints from the perturbative unitarity and the vacuum stability conditions.
We have found that the smaller mass regime for the additional Higgs bosons is preferable.
For example, if we require the cutoff scale of this model to be above 10~TeV, 
$H^0$ should be lighter than 800 GeV in the case of $m_A^{}=m_{H^\pm}^{} = m_H^{}+90$ GeV
and $\sin(\beta-\alpha) = 1$. 
Another consequence of the large Yukawa coupling is multi-muon final states at the LHC.
We have found that the region with $m_H^{} \lesssim 640$ GeV is excluded at 95\% CL
by the LHC data with 13 TeV of the collision energy and 35.9 fb$^{-1}$ of the integrated luminosity.
From these constraints,
we conclude that the cutoff scale of the $\mu$THDM is higher than 10~TeV but have to be lower than 100~TeV
if the model solves the muon $g-2$ anomaly within 1$\sigma$ level.

At the end, we briefly discuss how to weaken the constraint from the multi-muon signature at the LHC and make the 
cutoff scale higher.
One possible way is to add neutral and stable particles which couple to the additional Higgs bosons.
Then new decay modes of the additional Higgs bosons can open and the rate of the multi-muon final state can be reduced.
Another way is to embed this model into the context of composite THDMs~\cite{1105.5403,1612.05125} whose typical cutoff scale is
around 10~TeV. In that case, the model should be emerged from (unknown) UV dynamics.

\section*{Acknowledgments}
We would like to thank Howard E.~Haber and Pedro Ferreira for their comments.
We also thank Mihoko M.~Nojiri and Michihisa Takeuchi for their comments on LHC phenomenology.
This work was supported by JSPS KAKENHI Grant Number 16K17715 [TA].

\appendix
\section{Neutrino mass and mixing}\label{app:neutrino}

The observation of the neutrino oscillation shows three flavors of neutrinos $\nu_e$, $\nu_\mu$, and $\nu_\tau$ are mixed by large angles.
However, the global symmetry in our setup might forbid the mixing $\nu_\mu$ with the other neutrinos. 
In this section, we discuss dimension five operators for the Majorana neutrino mass matrices 
to see if they respect some symmetries.

The dimension five operators are given as follows.
\begin{align}
&
- \frac{c_{11}^{ij}}{M_{11}} (\overline{(L_L^i)}  \tilde{H}_1) (\widetilde{H}_1^{T} (L_L^c)^j)
- \frac{c_{12}^{ij}}{M_{12}} (\overline{(L_L^i)}  \tilde{H}_1) (\widetilde{H}_2^{T} (L_L^c)^j) \\
&
- \frac{c_{21}^{ij}}{M_{21}} (\overline{(L_L^i)}  \tilde{H}_2) (\widetilde{H}_1^{T} (L_L^c)^j)
- \frac{c_{22}^{ij}}{M_{22}} (\overline{(L_L^i)}  \tilde{H}_2) (\widetilde{H}_2^{T} (L_L^c)^j) \\
& + (h.c.).
\label{eq:dim5}
\end{align}
The $Z_4$ symmetry restricts the structure of the coefficient matrices as follows.
\begin{align}
 c_{11} =
\begin{pmatrix}
 (c_{11})^{ee} & (c_{11})^{e\tau} & 0\\
 (c_{11})^{\tau e} & (c_{11})^{\tau \tau} & 0 \\
 0 & 0 & 0 \\
\end{pmatrix} 
, \quad
 c_{12} =
\begin{pmatrix}
 0 & 0 & 0 \\
 0 & 0 & 0 \\
 0 & 0 & (c_{12})^{\mu \mu} \\
\end{pmatrix} 
, \\
 c_{21} =
\begin{pmatrix}
 0 & 0 & 0 \\
 0 & 0 & 0 \\
 0 & 0 & (c_{21})^{\mu \mu} \\
\end{pmatrix} 
, \quad
 c_{22} =
\begin{pmatrix}
 (c_{22})^{ee} & (c_{22})^{e\tau} & 0\\
 (c_{22})^{\tau e} & (c_{22})^{\tau \tau} & 0 \\
 0 & 0 & 0 \\
\end{pmatrix} 
.
\end{align}
From these matrices, we obtain the block diagonalized neutrino mass matrix, and thus the PMNS matrix is also block diagonalized.
This is inconsistent with the large mixing angle between $\nu_\mu$ and $\nu_{e,\tau}$. 
To obtain a realistic neutrino mass matrix, 
we add an SU(2) triplet scalar with $Y = -1$ ($\Delta$) which transforms under the $Z_4$ symmetry as
$\Delta \to - i \Delta.$
Using $\Delta$, we obtain following terms,
\begin{align}
 - c_\Delta^{ij} \bar{L}_L^i \Delta (L_L^c)^j,
\end{align}
where
\begin{align}
 c_{\Delta} =
\begin{pmatrix}
 0 & 0 & (c_{\Delta})^{e\mu}\\
0 & 0 & (c_{\Delta})^{\tau\mu}\\
 (c_{\Delta})^{\mu e} & (c_{\Delta})^{\mu \tau} & 0 
\label{eq:type-II}
\end{pmatrix} 
.
\end{align}
$\Delta$ obtains its VEV because of the coupling with the Higgs field via the following softly $Z_4$ breaking interactions:
\begin{align}
{\cal L} = \kappa_{11} \Delta H_1 H_1  + \kappa_{12} \Delta H_1 H_2  + \kappa_{22} \Delta H_2 H_2 + (h.c.). \label{eq:DeltaHH}
\end{align}
Using Eqs.~\eqref{eq:dim5}, \eqref{eq:type-II}, and \eqref{eq:DeltaHH},
we can obtain the neutrino mass matrix generated that does not contain zero-components,
\begin{align}
 m_{\nu} =&
\begin{pmatrix}
 (m_{\nu})^{e e} & (m_{\nu})^{e \tau} &  (m_{\nu})^{e \mu}  \\
 (m_{\nu})^{\tau e} & (m_{\nu})^{\tau \tau} &  (m_{\nu})^{\tau \mu} \\
  (m_{\nu})^{\mu e} &  (m_{\nu})^{\mu \tau} & (m_{\nu})^{\mu \mu} 
\label{eq:neutrino_mass}
\end{pmatrix}
.
\end{align}
It is possible to obtain realistic neutrino masses and the PMNS matrix from Eq.~\eqref{eq:neutrino_mass} without hard breaking of $Z_4$ symmetry.
We do not further discuss the neutrino physics in this paper.
As long as all the particles that arise from $\Delta$ are much heavier than all the other particles,
they are irrelevant with the phenomenology at the collider experiments.
In this sense, an extension which is discussed here does not affect to our analysis in the main part of this paper.

\section{Other discrete symmetries for $\mu$THDM}
\label{sec:Zn}

We briefly discuss other realizations of the $\mu$THDM. 
We assume a $Z_N$ symmetry to avoid FCNCs at the tree level.
It might be possible to use the other discrete symmetries for the realization of the model, but it is beyond the scope here.

Similar to the $Z_4$ symmetry discussed in the main part of this paper,
we assign non-trivial $Z_N$ charges to $\ell_L^\mu$, $\mu_R$, and $H_1$. 
All the other fields are singlet under the $Z_N$ symmetry.
The charge assignment is summarized in Table~\ref{tab:matter2}, where $a$, $b$, and $c$ are integers,
$a, b, c = 0, 1, 2, \cdots N-1$.
\begin{table}
\caption{The matter contents and the charge assignments. Here $\omega = \exp(2\pi i/N) $.}
\label{tab:matter2}
\begin{tabular}{|c||c|c|c||c|c|c||c|c|c||c|c|}
\hline
          & $q_L^{j}$ & $u_R^j$ & $d_R^j$ & $\ell_L^e$ & $\ell_L^\tau$ & $\ell_L^\mu$ & $e_R$ & $\tau_R$ & $\mu_R$    & $H_1$ & $H_2$ \\ \hline
SU(3)$_c$ & 3         & 3       & 3       & 1          & 1             & 1            & 1     & 1        & 1          & 1     & 1     \\ 
SU(2)$_L$ & 2         & 1       & 1       & 2          & 2             & 2            & 1     & 1        & 1          & 2     & 2     \\ 
 U(1)$_Y$ & 1/6       & 2/3     & $-1/3$  & $-1/2$     & $-1/2$        & $-1/2$       & $-1$  & $-1$     & $-1$       & 1/2   & 1/2   \\ 
    $Z_N$ & 1         & 1       & 1       & 1          & 1             & $\omega^a$   & 1     & 1        & $\omega^b$ & $\omega^c$   & 1     \\ \hline
\end{tabular}
\end{table}
The $Z_N$ charges have to satisfy the following conditions 
in order to obtain the muon specific texture for the lepton Yukawa matrices given in Eq.~\eqref{eq:texture_of_leptonYukawa}.
\begin{align}
& a \neq 0, \quad b \neq 0, \quad c \neq 0, \\
& -a + c \neq 0, \quad b + c \neq 0, \\
& -a +b +c = 0.
\end{align}
These conditions requires $N \geq 3$.

The $Z_N$ symmetry with the above conditions forbids $(H_1^\dagger H_1)(H_1^\dagger H_2)$ and $(H_2^\dagger H_2)(H_2^\dagger H_1)$.
Therefore the Higgs potential is given by Eq.~\eqref{eq:HiggsPotential}.
The $Z_N$ symmetry can also forbid the $\lambda_5$ term in the Higgs potential, $(H_1^\dagger H_2)^2$, if $\omega^{2c} \neq 1$.

Let us here discuss what happens if $\lambda_5 = 0$. 
In this case, the masses of $H^0$ and $A^0$ are degenerate in the large $\tan\beta$ limit.  
This can be understood by noting the appearance of an 
accidental global U(1) symmetry in the Higgs potential, which is similar to the Peccei-Quinn symmetry. 
Namely, the absence of the $\lambda_5$ and $m_3^2$ terms, the latter happens due to the large $\tan\beta$ limit under a fixed value of $M^2$ (see Eq.~(\ref{eq:def_M})), 
makes the Higgs potential invariant under the transformation, $H_{1,2} \to \exp(i \theta_{1,2}) H_{1,2}$.
This symmetry forces the CP-even neutral scalar to have the degenerate mass with the CP-odd neutral scalar.

This mass degeneracy reduces the contribution to the muon $g-2$, because the $A^0$ and $H^0$ loop effects are destructive. 
In order to compensate this reduction, we need to take smaller masses of $A^0$ and $H^0$. 
However, smaller masses are highly disfavored by the searches of multi-lepton final state at the LHC as discussed in Sec.~\ref{sec:LHC}. 
This is the reason why we choose the case with $\lambda_5 \neq 0$ which is realized by $\omega^{2c} = 1$ as mentioned above, and 
the $Z_4$ symmetry corresponds to the minimal choice for the realization of non-zero $\lambda_5$.

\section{Details on the constraints from the electroweak precision measurements}
\label{sec:EWPTapp}

We choose $\alpha_{\text{em}}$, $m_Z$, and $\sqrt{2} G_F$ as the input parameters. They relate to the model parameters as follows. 
\begin{align}
 4 \pi \alpha_{\text{em}} =& \frac{1}{1 - \frac{d \Pi_{\gamma \gamma}}{dq^2}(m_Z^2)} \left(\frac{1}{g^2} + \frac{1}{g'^2} \right)^{-2}, \\
 m_Z^2 =& \frac{g^2 + g'^2}{4} v^2 + \Pi_{ZZ}(m_Z^2), \\
 \sqrt{2} G_F =&  
  \frac{1}{v^2} \left( 1 + \frac{\delta g_W^{\mu}}{g}  - \frac{\Pi_{WW}(0)}{\frac{g^2}{4} v^2} \right),
\label{eq:GFdef}
\end{align}
where $v^2 = v_1^2 + v_2^2$, $g$ is the SU(2)$_L$ gauge coupling, $g'$ is the U(1)$_Y$ gauge coupling,
$\Pi_{ij}$'s are the gauge boson self-energies, 
and $\delta g_{W}^{\mu}$ is the sum of the vertex corrections to $W$-$\mu$-$\nu_\mu$ coupling 
at zero momentum with the wave function renormalization effects. 
The Fermi constant receives the non-negligible effect from the vertex correction as can be seen Eq.~\eqref{eq:GFdef}.
Therefore the vertex correction $\delta g_W^{\mu}$ affects every observables through the replacement of $v^2$ by $G_F$.
This effect is a reason why we cannot use the $S$, $T$ and $U$ parameters directly.

We derive the deviations of the model prediction from the SM prediction in the same manner as in \cite{
PRLTA.65.964,PHRVA.D46.381},
$\Delta O \equiv {\cal O}_{\text{model}} - {\cal O}_{\text SM}$. 
The result is complicated but summarized by the following modified version of the $S$, $T$ and $U$ parameters.
\begin{align}
 \alpha_{\text{em}} \tilde{T} =& \left( \frac{\Pi_{WW}(0)}{m_W^2} - \frac{\delta g_W^{\mu}}{g} \right)- \frac{\Pi_{ZZ}(0)}{m_Z^2}, \\
 \frac{\alpha_{\text{em}}}{4 s_0^2 c_0^2} \tilde{S} =&
\frac{\Pi_{ZZ}(m_Z^2) - \Pi_{ZZ}(0)}{m_Z^2} - \frac{c_0^2 - s_0^2}{c_0 s_0} \frac{\Pi_{Z\gamma}(m_Z^2)}{m_Z^2} - \frac{\Pi_{\gamma \gamma}(m_Z^2)}{m_Z^2},\\
 \frac{\alpha_{\text{em}}}{4 s_0^2} (\tilde{S}+\tilde{U}) =&
\frac{\Pi_{WW}(m_W^2) - \Pi_{WW}(0)}{m_W^2} + \frac{\delta g_W^{\mu}}{g} - \frac{c_0}{s_0} \frac{\Pi_{Z\gamma}(m_Z^2)}{m_Z^2} - \frac{\Pi_{\gamma \gamma}(m_Z^2)}{m_Z^2},
\\
\alpha_{\text{em}} \tilde{W}
=&
\frac{d \Pi_{WW}}{dq^2}(m_W^2)
-
\frac{\Pi_{WW}(m_W^2) - \Pi_{WW}(0)}{m_W^2}
- \frac{\delta g_W^{\mu}}{g_W}
,\\
\alpha_{\text{em}} \tilde{Z}
=&
 \frac{d \Pi_{ZZ}}{d q^2}(m_Z^2)
- \frac{\Pi_{ZZ}(m_Z^2) - \Pi_{ZZ}(0)}{m_Z^2} 
.
\end{align}
We also use $s_0^2$ and $c_0^2 = 1 - s_0^2$ that are defined by the input values as
\begin{align}
 s_0^2 c_0^2 = \frac{\alpha_{\text{em}} \pi}{ m_Z^2 \sqrt{2} G_F}.
\label{eq:def_s0}
\end{align}
$\tilde{S}$ is the same as $S$ defined in PDG~\cite{Olive:2016xmw}.
If $\delta g_W^{\mu}/g = 0$, then , $\tilde{T}$ and $\tilde{U}$ becomes the same as $T$ and $U$ given in PDG, respectively.
$\tilde{W}$ and $\tilde{Z}$ are negligible if new particles are much heavier than the electroweak gauge bosons. 
They cannot be ignored in our setup.
In $\delta g_W^{\mu}/g \to 0$ limit, 
$\tilde{W}$ and $\tilde{Z}$ becomes $W$ and $V$ defined in \cite{hep-ph/9306267, hep-ph/9307337}, respectively.
Using these parameters, we find the following expressions for $\Delta O \equiv {\cal O}_{\text{model}} - {\cal O}_{\text SM}$.
\begin{align}
 \Delta m_W
=& 
\frac{1}{4} \frac{1}{s_0} \left(\frac{4\pi \alpha_{\text{em}}}{\sqrt{2} G_F} \right)^{1/2}
\Biggl[
- \frac{\alpha_{\text{em}}}{2 (c_0^2 - s_0^2)} \tilde{S}  
+ \frac{c_0^2}{c_0^2 - s_0^2} \alpha_{\text{em}} \tilde{T}
+ \frac{\alpha_{\text{em}}}{4 s_0^2} \tilde{U}
\Biggr]
,\\
 \Delta \Gamma(W \to f f')
=&
\frac{m_W N_c}{12}
\frac{\alpha_{\text{em}}}{s_0^2}
\Bigg[
- \frac{1}{2 (c_0^2 - s_0^2)} \alpha_{\text{em}} \tilde{S}
+ \frac{c_0^2}{c_0^2 - s_0^2} \alpha_{\text{em}} \tilde{T}
+ \frac{\alpha_{\text{em}}}{4 s_0^2}  \tilde{U}\notag\\
&\hspace{2.5cm}+ \alpha_{\text{em}} \tilde{W}
+ 2 \frac{\delta g_{Wff'}(m_W)}{g_W}
\Bigg]
,\\
\Delta \Gamma(Z \to ff')
=&
 \frac{m_Z}{24} N_c \Biggl[
 -Q (j_3 - 2 s_0^2 Q) 
  \frac{\alpha_{\text{em}} \tilde{S}}{2 (c_0^2 - s_0^2)} 
\nonumber\\
& \qquad \qquad
+
\left(
\left( ( j_3 - s_0^2 Q )^2  +  (- s_0^2 Q )^2 \right)
 +2 Q (j_3 - 2 s_0^2 Q)  \frac{s_0^2 c_0^2}{c_0^2 - s_0^2} 
\right)
\alpha_{\text{em}} \tilde{T}
\nonumber\\
& \qquad \qquad
+
\left( ( j_3 - s_0^2 Q )^2  +  (- s_0^2 Q )^2 \right) \alpha_{\text{em}} \tilde{Z}
\nonumber\\
& \qquad \qquad
+ 2 (j_3 - s_0^2 Q) \left(  \frac{\delta g_{Zff'}^L}{g_Z} \right)
- 2 s_0^2 Q \left( \frac{\delta g_{Zff'}^R}{g_Z} \right)
\Biggr]
,\\
 \Delta A_f
=&
\frac{4 s_0^2 Q (j_3 - s_0^2 Q)}{[(j_3 - s_0^2 Q)^2 + (s_0^2 Q)^2]^2}
\Bigg[
- j_3 Q \frac{\alpha_{\text{em}} \tilde{S}}{4 (c_0^2 - s_0^2)}
+ j_3 Q \frac{s_0^2 c_0^2}{c_0^2 - s_0^2} \alpha_{\text{em}} \tilde{T}\notag\\
&\hspace{4.5cm}+ \frac{\delta g_{Zff'}^{L}}{g_Z}
+ \frac{\delta g_{Zff'}^{R}}{g_Z}
\Bigg]
,\\
 \Delta A_{FB}^{0,f}=& \frac{3}{4} \left( \Delta A_f A_e^{\text{SM}} + A_f^{\text{SM}} \Delta A_e \right)
,\\
 \Delta R_{\ell}
\simeq&
 R_{\ell}^{\text{SM}}
\left(
\frac{\Delta \Gamma(Z \to had)}{\left.\Gamma(Z \to had)\right.|_{\text{SM}}}
-
\frac{\Delta \Gamma(Z \to \ell \ell)}{\left.\Gamma(Z \to \ell \ell)\right.|_{\text{SM}}}
\right)
,\\ 
 \Delta R_{q}
\simeq&
 R_{q}^{\text{SM}}
\left(
\frac{\Delta \Gamma(Z \to qq)}{\left.\Gamma(Z \to qq)\right.|_{\text{SM}}}
-
\frac{\Delta \Gamma(Z \to had)}{\left.\Gamma(Z \to had)\right.|_{\text{SM}}}
\right)
,
\end{align}
where $j_3$ and $Q$ are isospin and electric charge of external fermions, resepctively. $N_c=3$ (for external quarks) or 1 (for external leptons). 
We also introduced the following quantities:
\begin{align}
 g_W =& \left( \frac{4 \pi \alpha_{\text{em}}}{s_0^2} \right)^{1/2}, \\
 g_Z =& \left( \frac{4 \pi \alpha_{\text{em}}}{s_0^2 c_0^2} \right)^{1/2}, \\
 \Gamma(Z \to had) =& \sum_{q} \Gamma(Z \to qq),
\end{align}
and $\delta g_{Z ff'}^{L, R}$ are calculated at $q^2 = m_Z^2$.
$\delta g_{Wff'}(m_W)$ and $\delta g_{Z ff'}^{L, R}$ are only relevant for the muon sector and negligible in the other sector.

We use the values given by PDG~\cite{Olive:2016xmw}.
Input parameters are
\begin{align}
& G_F = 1.1663787 10^{-5}~\text{GeV}^{-2}, \quad m_Z = 91.1876~\text{GeV}, \quad \alpha_{\text{em}}^{-1}(m_Z) = 127.950, \\
& m_{\mu} = 0.1056583745~\text{GeV}, \quad 
\end{align}
The SM predictions and the values to be fitted are given in Table~\ref{tab:data}.
\begin{table}[htb]
\caption{The electroweak precision data given by PDG~\cite{Olive:2016xmw}.}
 \begin{tabular}{|c|c|c|} 
\hline
Quantity & Value & SM \\ \hline \hline
$m_W$[GeV] & 80.385 $\pm$ 0.015 & 80.361 $\pm$ 0.006 \\ 
$\Gamma_W$[GeV] & 2.085 $\pm$ 0.042 & 2.089 $\pm$ 0.001 \\ 
$\Gamma_Z$[GeV] & 2.4952 $\pm$ 0.0023 & 2.4943 $\pm$ 0.0008 \\ 
$\Gamma$(had)[GeV] & 1.7444 $\pm$ 0.0020 & 1.7420 $\pm$ 0.0008 \\ 
$\Gamma$(inv)[MeV] & 499.0 $\pm$ 1.5 & 501.66 $\pm$ 0.05 \\ 
$\Gamma(\ell^{+} \ell^{-})$[MeV] & 83.984 $\pm$ 0.086 & 83.995 $\pm$ 0.010 \\ 
$\Gamma(\mu \mu)$[MeV] & 83.99 $\pm$ 0.18 & 83.995 $\pm$ 0.010 \\ 
$R_e$ & 20.804 $\pm$ 0.050 & 20.734 $\pm$ 0.010 \\ 
$R_\mu$ & 20.785 $\pm$ 0.033 & 20.734 $\pm$ 0.010 \\ 
$R_\tau$ & 20.764 $\pm$ 0.045 & 20.779 $\pm$ 0.010 \\ 
$R_b$ & 0.21629 $\pm$ 0.00066 & 0.21579 $\pm$ 0.00003 \\ 
$R_c$ & 0.1721 $\pm$ 0.0030 & 0.17221 $\pm$ 0.00003 \\ 
$A^{(0,e)}_{FB}$ & 0.0145 $\pm$ 0.0025 & 0.01622 $\pm$ 0.00009 \\ 
$A^{(0,\mu)}_{FB}$ & 0.0169 $\pm$ 0.0013 & 0.01622 $\pm$ 0.00009 \\ 
$A^{(0,\tau)}_{FB}$ & 0.0188 $\pm$ 0.0017 & 0.01622 $\pm$ 0.00009\\ 
$A^{(0,b)}_{FB}$ & 0.0992 $\pm$ 0.0016 & 0.1031 $\pm$ 0.0003 \\ 
$A^{(0,c)}_{FB}$ & 0.0707 $\pm$ 0.0035 & 0.0736 $\pm$ 0.0002 \\ 
$A^{(0,s)}_{FB}$ & 0.0876 $\pm$ 0.0114 & 0.1032 $\pm$ 0.0003 \\ 
$A_e$    & 0.1515 $\pm$ 0.0019 & 0.1470 $\pm$ 0.0004 \\ 
$A_\mu$  & 0.142  $\pm$ 0.015  & 0.1470 $\pm$ 0.0004 \\ 
$A_\tau$ & 0.143  $\pm$ 0.004  & 0.1470 $\pm$ 0.0004 \\ 
$A_b$ & 0.923 $\pm$ 0.020 & 0.9347 \\ 
$A_c$ & 0.670 $\pm$ 0.027 & 0.6678 $\pm$ 0.0002 \\ 
$A_s$ & 0.90  $\pm$ 0.09  & 0.9356 \\ 
\hline
\end{tabular}
\label{tab:data}
\end{table}
We construct likelihood function,
\begin{align}
 \chi^2 = \sum_O \left( \frac{\Delta O - (O_{\text{obs}} - O_{\text{SM}})}{\sigma_{\text{obs}}} \right)^2,
\end{align}
and perform the $\chi^2$ analysis.

We apply the above formula to the $\mu$THDM.
For $\sin(\beta - \alpha) = 1$, we find 
{\allowdisplaybreaks 
\begin{align}
 \Pi_{WW}(p^2)
=&
- \frac{g_W^2}{4 (4 \pi)^2}
\Big[ A_0(m_A^2)  +  A_0(m_H^2) + 2  A_0(m_{H^{\pm}}^2)  \notag\\
& \hspace{1.5cm}- 4 B_{00}(p^2, m_H^2, m_{H^{\pm}}^2) - 4 B_{00}(p^2, m_A^2, m_{H^{\pm}}^2) \Big]
,\\
 \Pi_{ZZ}(p^2)
=&
- \frac{g_Z^2}{4 (4 \pi)^2}
\Biggl[
 A_0(m_A^2)  +  A_0(m_H^2) + 2 (c_0^2 - s_0^2)^2 A_0(m_{H^{\pm}}^2) 
\nonumber\\
&
\qquad \qquad \qquad
- 4 B_{00}(p^2, m_H^2, m_A^2) - 4 (c_0^2 - s_0^2)^2 B_{00}(p^2, m_{H^{\pm}}^2, m_{H^{\pm}}^2)
\Biggr],\\
 \Pi_{\gamma \gamma}(p^2)
=&
\frac{e^2}{(4 \pi)^2}
\Big[  -2 A_0(m_{H^{\pm}}^2) + 4 B_{00}(p^2, m_{H^{\pm}}^2, m_{H^{\pm}}^2) \Big]
,\\
 \Pi_{Z \gamma}(p^2)
=&
\frac{c_0^2 - s_0^2}{2 s_0c_0}  \Pi_{\gamma \gamma}(p^2),\\
 \delta g_{Z\mu\mu}^L
\simeq&
\frac{g_Z}{(4\pi)^2}
\frac{m_{\mu}^2}{v^2}
\tan^2\beta \notag\\
&\times
\Biggl(
- \frac{1}{2}
  \Bigl[ B_{1}(_{\mu,\mu,a}) + B_1(_{\mu,\mu,H}) + 4 C_{00}(_{Z,\mu,\mu,H,A,\mu}) \Bigr]
\nonumber\\
&
\qquad
+ s_0^2 
\Bigl[
 -1 + B_1(_{\mu,\mu,A}) + B_1(_{\mu,\mu,H}) + 2C_{00}(_{Z,\mu,\mu,\mu,\mu,A}) + 2C_{00}(_{Z,\mu,\mu,\mu,\mu,H})
\nonumber\\
& \qquad \qquad
 - m_Z^2 C_{12}(_{\mu,Z,\mu,A,\mu,\mu})  - m_Z^2 C_{12}(_{\mu,Z,\mu,H,\mu,\mu})
\Bigr]
\Biggr)
,\\ 
 \delta g_{Z\mu\mu}^R
\simeq&
\frac{g_Z}{(4\pi)^2}
\frac{m_{\mu}^2}{v^2}
\tan^2\beta
\nonumber\\
&\times
\Biggl(
  -C_{00}(_{Z,\mu,\mu,\mu,\mu,A}) -C_{00}(_{Z,\mu,\mu,\mu,\mu,H}) + 2 C_{00}(_{Z,\mu,\mu,0,0,H^{\pm}})
\nonumber\\
&\qquad
  +2 C_{00}(_{Z,\mu,\mu,H,A,\mu}) -2 C_{00}(_{Z,\mu,\mu,H^{\pm},H^{\pm},0})
\nonumber\\
&\qquad
+ \frac{1}{2} m_Z^2 
\left[ C_{12}(_{\mu,Z,\mu,A,\mu,\mu}) + C_{12}(_{\mu,Z,\mu,H,\mu,\mu}) - 2 C_{12}(_{\mu,Z,\mu,H^{\pm},0,0}) \right]
\nonumber\\
&\qquad
+ s_0^2
\Biggl[
-1 + B_{1}(_{\mu,\mu,A^0}) + B_{1}(_{\mu,\mu,H}) + 2 C_{00}(_{Z,\mu,\mu,\mu,\mu,A}) +2 C_{00}(_{Z,\mu,\mu,\mu,\mu,H})
\nonumber\\
&\qquad \qquad
+ 2 B_{1}(_{\mu,0,H^{\pm}}) 
+4 C_{00}(_{Z,\mu,\mu,H^{\pm},H^{\pm},0})
\nonumber\\
&\qquad \qquad
-m_Z^2 C_{12}(_{\mu,Z,\mu,A,\mu,\mu})
-m_Z^2 C_{12}(_{\mu,Z,\mu,H,\mu,\mu})
\Biggr]
\Biggr)
,
\end{align}
\newpage
\begin{align}
\delta g_{Z\nu_{\mu}\nu_{\mu}}^L
\simeq&
\frac{g_Z}{(4\pi)^2}
\frac{m_{\mu}^2}{v^2}
\tan^2\beta
\nonumber\\
&\times
\Biggl(
B_1(_{0,\mu,H^{\pm}}) + 2 C_{00}(_{Z,0,0,H^{\pm},H^{\pm},\mu})
\nonumber\\
&
\quad
+s_0^2 
 \Bigl[
   4 C_{00}(_{0,Z,0,H^{\pm},\mu,\mu}) -4 C_{00}(_{Z,0,0,H^{\pm},H^{\pm},\mu}) - 2 m_Z^2 C_{12}(_{0,Z,0,H^{\pm},\mu,\mu})-1
 \Bigr]
\Biggr)
,\\ 
\delta g_{W \mu \nu_{\mu}}^L(m_W^2)
\simeq&
\frac{g_W}{(4\pi)^2}
\frac{m_{\mu}^2}{v^2}
\tan^2\beta
\nonumber\\
&\times
\Biggl(
- \frac{1}{2} - \frac{1}{4} B_0(_{0,A,A}) - \frac{1}{4} B_0(_{0,H,H}) - \frac{1}{2} B_0(_{0,H^{\pm},H^{\pm}})\notag\\
&\hspace{1cm}+ 2 C_{00}(_{0,W,0,0,A,H^{\pm}}) + 2 C_{00}(_{0,W,0,0,H,H^{\pm}})
\Biggr)
,\\ 
\delta g_{W}^{\mu}
\simeq&
\frac{g_W}{(4\pi)^2}
\frac{m_{\mu}^2}{v^2}
\tan^2\beta
\Biggl(
1 
- \frac{m_A^2 + m_{H^{\pm}}^2}{4 (m_A^2 - m_{H^{\pm}}^2)} \ln \frac{m_A^2}{m_{H^{\pm}}^2}
- \frac{m_H^2 + m_{H^{\pm}}^2}{4 (m_H^2 - m_{H^{\pm}}^2)} \ln \frac{m_H^2}{m_{H^{\pm}}^2}
\Biggr)
.
\end{align}
}
where
\begin{align}
 (_{a,b,c,\cdots}) =& (m_a^2, m_b^2, m_c^2, \cdots), \\
 (_{\cdots,0,\cdots}) =& (\cdots, 0, \cdots).
\end{align}
The notation of $A$, $B$, and $C$ function is the same as the notation used by \texttt{LoopTools}~\cite{hep-ph/9807565}.
\begin{align}
A_0(m^2)
=&
m^2 \left( \frac{1}{\bar{\epsilon}}  + 1 + \ln \frac{\mu^2}{m^2} \right),
\\
B_0(q^2,m_1^2,m_2^2)
=&
\frac{1}{\bar{\epsilon}} 
+ \int_0^1 dx \ln \frac{\mu^2}{m_1^2 x + m_2^2 (1-x) - q^2 x (1-x) },
\\
B_1(q^2,m_1^2,m_2^2)
=&
- \frac{1}{2 \bar{\epsilon}} 
- \int_0^1 dx (1-x) \ln \frac{\mu^2}{m_1^2 x + m_2^2 (1-x) - q^2 x (1-x) },
\\
 B_{00}(q^2,m_1^2,m_2^2)
 =&
 \frac{1}{4} \left( \frac{1}{\bar{\epsilon}} + 1 \right) \left[ m_1^2 + m_2^2 - \frac{1}{3} q^2\right]
 \nonumber\\
 &
 + \frac{1}{2}
  \int_0^1 dx (m_1^2 x + m_2^2 (1-x) - q^2 x (1-x)) \notag\\
&\hspace{2cm}\times \ln \frac{\mu^2}{m_1^2 x + m_2^2 (1-x) - q^2 x (1-x) },
\\
C_{00}(p_1^2,p_2^2,p_3^2,m_1^2,m_2^2,m_3^2)
=&
\frac{1}{4\bar{\epsilon}}
+
\frac{1}{2}
\int_{xyz}
\ln \frac{\mu^2}{m_1^2 x + m_2^2 y + m_3^2 z - p_1^2 xy -p_2^2 yz - p_3^2 zx}
,
\\
C_{12}(p_1^2,p_2^2,p_3^2,m_1^2,m_2^2,m_3^2)
=&
-\int_{xyz} \frac{yz}{m_1^2 x + m_2^2 y + m_3^2 z - p_1^2 xy -p_2^2 yz - p_3^2 zx}.
\end{align}
where
\begin{align}
 \int_{xyz} = \int dx dy dz \delta(x+y+z-1) = \int_0^1 dx \int_0^{1-x}dy 
\end{align}

We vary the four model parameters ($m_H$, $m_A$, $m_{H^{\pm}}$, $t_\beta$) and try to fit
the 24 observables in Table.~\ref{tab:data}.
We find that the minimum value of $\chi^2$ is given by $\chi_{\text{min.}}^2 = 23.7587$ at 
($m_H$, $m_A$, $m_{H^{\pm}}$, $t_\beta$) = (59.4~GeV, 398~GeV, 402~GeV, 686).
We calculate $\Delta \chi^2 \equiv \chi^2 - \chi_{\text{min.}}^2$ 
by varying $m_H$ and $t_\beta$ with fixed values for $m_A$ and $m_{H^{\pm}}$.
The result is shown in Fig.~\ref{fig:g-2_and_constraints} where the gray region is excluded at 95\% CL.



\begin{thebibliography}{99}
\bibitem{Bennett:2006fi} 
  G.~W.~Bennett {\it et al.} [Muon g-2 Collaboration],
  Phys.\ Rev.\ D {\bf 73}, 072003 (2006)
  doi:10.1103/PhysRevD.73.072003
  [hep-ex/0602035].


\bibitem{Davier:2010nc} 
  M.~Davier, A.~Hoecker, B.~Malaescu and Z.~Zhang,
  Eur.\ Phys.\ J.\ C {\bf 71}, 1515 (2011)
  Erratum: [Eur.\ Phys.\ J.\ C {\bf 72}, 1874 (2012)]
  doi:10.1140/epjc/s10052-012-1874-8, 10.1140/epjc/s10052-010-1515-z
  [arXiv:1010.4180 [hep-ph]].


\bibitem{Hagiwara:2011af} 
  K.~Hagiwara, R.~Liao, A.~D.~Martin, D.~Nomura and T.~Teubner,
  J.\ Phys.\ G {\bf 38}, 085003 (2011)
  doi:10.1088/0954-3899/38/8/085003
  [arXiv:1105.3149 [hep-ph]].


\bibitem{0902.3360} 
  F.~Jegerlehner and A.~Nyffeler,
  Phys.\ Rept.\  {\bf 477}, 1 (2009)
  doi:10.1016/j.physrep.2009.04.003
  [arXiv:0902.3360 [hep-ph]].


\bibitem{Grange:2015fou} 
  J.~Grange {\it et al.} [Muon g-2 Collaboration],
  arXiv:1501.06858 [physics.ins-det].


\bibitem{Iinuma:2011zz} 
  H.~Iinuma [J-PARC muon g-2/EDM Collaboration],
  J.\ Phys.\ Conf.\ Ser.\  {\bf 295}, 012032 (2011).
  doi:10.1088/1742-6596/295/1/012032

\bibitem{1409.3199} 
  A.~Broggio, E.~J.~Chun, M.~Passera, K.~M.~Patel and S.~K.~Vempati,
  JHEP {\bf 1411}, 058 (2014)
  doi:10.1007/JHEP11(2014)058
  [arXiv:1409.3199 [hep-ph]].


\bibitem{1412.4874} 
  L.~Wang and X.~F.~Han,
  JHEP {\bf 1505}, 039 (2015)
  doi:10.1007/JHEP05(2015)039
  [arXiv:1412.4874 [hep-ph]].


\bibitem{Barger:1989fj} 
  V.~D.~Barger, J.~L.~Hewett and R.~J.~N.~Phillips,
  Phys.\ Rev.\ D {\bf 41}, 3421 (1990).
  doi:10.1103/PhysRevD.41.3421


\bibitem{hep-ph/9401311} 
  Y.~Grossman,
  Nucl.\ Phys.\ B {\bf 426}, 355 (1994)
  doi:10.1016/0550-3213(94)90316-6
  [hep-ph/9401311].


\bibitem{0902.4665} 
  M.~Aoki, S.~Kanemura, K.~Tsumura and K.~Yagyu,
  Phys.\ Rev.\ D {\bf 80}, 015017 (2009)
  doi:10.1103/PhysRevD.80.015017
  [arXiv:0902.4665 [hep-ph]].


\bibitem{Glashow:1976nt} 
  S.~L.~Glashow and S.~Weinberg,
  Phys.\ Rev.\ D {\bf 15}, 1958 (1977).
  doi:10.1103/PhysRevD.15.1958




\bibitem{1502.07824} 
  Y.~Omura, E.~Senaha and K.~Tobe,
  JHEP {\bf 1505}, 028 (2015)
  doi:10.1007/JHEP05(2015)028
  [arXiv:1502.07824 [hep-ph]].


\bibitem{1511.05162} 
  T.~Han, S.~K.~Kang and J.~Sayre,
  JHEP {\bf 1602}, 097 (2016)
  doi:10.1007/JHEP02(2016)097
  [arXiv:1511.05162 [hep-ph]].


\bibitem{1511.08880} 
  Y.~Omura, E.~Senaha and K.~Tobe,
  Phys.\ Rev.\ D {\bf 94}, no. 5, 055019 (2016)
  doi:10.1103/PhysRevD.94.055019
  [arXiv:1511.08880 [hep-ph]].


\bibitem{1504.07059} 
  T.~Abe, R.~Sato and K.~Yagyu,
  JHEP {\bf 1507}, 064 (2015)
  doi:10.1007/JHEP07(2015)064
  [arXiv:1504.07059 [hep-ph]].


\bibitem{1605.06298} 
  E.~J.~Chun and J.~Kim,
  JHEP {\bf 1607}, 110 (2016)
  doi:10.1007/JHEP07(2016)110
  [arXiv:1605.06298 [hep-ph]].

\bibitem{Gunion:2002zf} 
  J.~F.~Gunion and H.~E.~Haber,
  Phys.\ Rev.\ D {\bf 67}, 075019 (2003)
  doi:10.1103/PhysRevD.67.075019
  [hep-ph/0207010].
  

\bibitem{1606.02266} 
  G.~Aad {\it et al.} [ATLAS and CMS Collaborations],
  JHEP {\bf 1608}, 045 (2016)
  doi:10.1007/JHEP08(2016)045
  [arXiv:1606.02266 [hep-ex]].


\bibitem{Dedes:2001nx} 
  A.~Dedes and H.~E.~Haber,
  JHEP {\bf 0105}, 006 (2001)
  doi:10.1088/1126-6708/2001/05/006
  [hep-ph/0102297].

\bibitem{BZ1} 
  J.~D.~Bjorken and S.~Weinberg,
  Phys.\ Rev.\ Lett.\  {\bf 38}, 622 (1977);

\bibitem{BZ2}
  S.~M.~Barr and A.~Zee,
  Phys.\ Rev.\ Lett.\  {\bf 65}, 21 (1990)
  [Erratum-ibid.\  {\bf 65}, 2920 (1990)].

\bibitem{PHRVA.D18.2574} 
  N.~G.~Deshpande and E.~Ma,
  Phys.\ Rev.\ D {\bf 18}, 2574 (1978).
  doi:10.1103/PhysRevD.18.2574


\bibitem{Sher:1988mj} 
  M.~Sher,
  Phys.\ Rept.\  {\bf 179}, 273 (1989).
  doi:10.1016/0370-1573(89)90061-6


\bibitem{PHLTA.B449.89} 
  S.~Nie and M.~Sher,
  Phys.\ Lett.\ B {\bf 449}, 89 (1999)
  doi:10.1016/S0370-2693(99)00019-2
  [hep-ph/9811234].


\bibitem{PHLTA.B471.182} 
  S.~Kanemura, T.~Kasai and Y.~Okada,
  Phys.\ Lett.\ B {\bf 471}, 182 (1999)
  doi:10.1016/S0370-2693(99)01351-9
  [hep-ph/9903289].



\bibitem{PHLTA.B313.155} 
  S.~Kanemura, T.~Kubota and E.~Takasugi,
  Phys.\ Lett.\ B {\bf 313}, 155 (1993)
  doi:10.1016/0370-2693(93)91205-2
  [hep-ph/9303263].


\bibitem{hep-ph/0006035} 
  A.~G.~Akeroyd, A.~Arhrib and E.~M.~Naimi,
  Phys.\ Lett.\ B {\bf 490}, 119 (2000)
  doi:10.1016/S0370-2693(00)00962-X
  [hep-ph/0006035].


\bibitem{PHRVA.D72.115010} 
  I.~F.~Ginzburg and I.~P.~Ivanov,
  Phys.\ Rev.\ D {\bf 72}, 115010 (2005)
  doi:10.1103/PhysRevD.72.115010
  [hep-ph/0508020].

\bibitem{Kanemura:2015ska} 
  S.~Kanemura and K.~Yagyu,
  Phys.\ Lett.\ B {\bf 751}, 289 (2015)
  doi:10.1016/j.physletb.2015.10.047
  [arXiv:1509.06060 [hep-ph]].


\bibitem{1309.7223} 
  F.~Staub,
  Comput.\ Phys.\ Commun.\  {\bf 185}, 1773 (2014)
  doi:10.1016/j.cpc.2014.02.018
  [arXiv:1309.7223 [hep-ph]].


\bibitem{PRLTA.65.964} 
  M.~E.~Peskin and T.~Takeuchi,
  Phys.\ Rev.\ Lett.\  {\bf 65}, 964 (1990).
  doi:10.1103/PhysRevLett.65.964


\bibitem{PHRVA.D46.381} 
  M.~E.~Peskin and T.~Takeuchi,
  Phys.\ Rev.\ D {\bf 46}, 381 (1992).
  doi:10.1103/PhysRevD.46.381


\bibitem{1108.2040} 
  C.~Degrande, C.~Duhr, B.~Fuks, D.~Grellscheid, O.~Mattelaer and T.~Reiter,
  Comput.\ Phys.\ Commun.\  {\bf 183}, 1201 (2012)
  doi:10.1016/j.cpc.2012.01.022
  [arXiv:1108.2040 [hep-ph]].


\bibitem{1310.1921} 
  A.~Alloul, N.~D.~Christensen, C.~Degrande, C.~Duhr and B.~Fuks,
  Comput.\ Phys.\ Commun.\  {\bf 185}, 2250 (2014)
  doi:10.1016/j.cpc.2014.04.012
  [arXiv:1310.1921 [hep-ph]].


\bibitem{1106.0522} 
  J.~Alwall, M.~Herquet, F.~Maltoni, O.~Mattelaer and T.~Stelzer,
  JHEP {\bf 1106}, 128 (2011)
  doi:10.1007/JHEP06(2011)128
  [arXiv:1106.0522 [hep-ph]].


\bibitem{CMS:2017wua} 
  CMS Collaboration [CMS Collaboration],
  CMS-PAS-EXO-17-006.


\bibitem{hep-ph/0603175} 
  T.~Sjostrand, S.~Mrenna and P.~Z.~Skands,
  JHEP {\bf 0605}, 026 (2006)
  doi:10.1088/1126-6708/2006/05/026
  [hep-ph/0603175].


\bibitem{1307.6346} 
  J.~de Favereau {\it et al.} [DELPHES 3 Collaboration],
  JHEP {\bf 1402}, 057 (2014)
  doi:10.1007/JHEP02(2014)057
  [arXiv:1307.6346 [hep-ex]].


\bibitem{hep-ex/9902006} 
  T.~Junk,
  Nucl.\ Instrum.\ Meth.\ A {\bf 434}, 435 (1999)
  doi:10.1016/S0168-9002(99)00498-2
  [hep-ex/9902006].


\bibitem{Read:2000ru} 
  A.~L.~Read,
  In *Geneva 2000, Confidence limits* 81-101


\bibitem{Read:2002hq} 
  A.~L.~Read,
  J.\ Phys.\ G {\bf 28}, 2693 (2002).
  doi:10.1088/0954-3899/28/10/313


\bibitem{1105.5403} 
  J.~Mrazek, A.~Pomarol, R.~Rattazzi, M.~Redi, J.~Serra and A.~Wulzer,
  Nucl.\ Phys.\ B {\bf 853}, 1 (2011)
  doi:10.1016/j.nuclphysb.2011.07.008
  [arXiv:1105.5403 [hep-ph]].


\bibitem{1612.05125} 
  S.~De Curtis, S.~Moretti, K.~Yagyu and E.~Yildirim,
  Phys.\ Rev.\ D {\bf 94}, no. 5, 055017 (2016)
  doi:10.1103/PhysRevD.94.055017
  [arXiv:1602.06437 [hep-ph]];

  S.~De Curtis, S.~Moretti, K.~Yagyu and E.~Yildirim,
  arXiv:1610.02687 [hep-ph].


\bibitem{Olive:2016xmw} 
  C.~Patrignani {\it et al.} [Particle Data Group],
  Chin.\ Phys.\ C {\bf 40}, no. 10, 100001 (2016).
  doi:10.1088/1674-1137/40/10/100001


\bibitem{hep-ph/9306267} 
  I.~Maksymyk, C.~P.~Burgess and D.~London,
  Phys.\ Rev.\ D {\bf 50}, 529 (1994)
  doi:10.1103/PhysRevD.50.529
  [hep-ph/9306267].


\bibitem{hep-ph/9307337} 
  C.~P.~Burgess, S.~Godfrey, H.~Konig, D.~London and I.~Maksymyk,
  Phys.\ Lett.\ B {\bf 326}, 276 (1994)
  doi:10.1016/0370-2693(94)91322-6
  [hep-ph/9307337].


\bibitem{hep-ph/9807565} 
  T.~Hahn and M.~Perez-Victoria,
  Comput.\ Phys.\ Commun.\  {\bf 118}, 153 (1999)
  doi:10.1016/S0010-4655(98)00173-8
  [hep-ph/9807565].




 
\end{thebibliography}
\end{document}